\documentclass[aps,prb,preprint,superscriptaddress]{revtex4}

\usepackage{amsmath}
\usepackage{graphicx}
\usepackage{natmove}

\usepackage{color}
\usepackage[normalem]{ulem}
\usepackage{times}

\begin{document}

\title{
When are rough surfaces sticky?
}

\author{Lars Pastewka}
\affiliation{Johns Hopkins University, Department of Physics and Astronomy,
  3400 North Charles Street, Baltimore, MD 21218, USA}
\affiliation{Fraunhofer IWM, MikroTribologie Centrum $\mu$TC, W\"ohlerstra\ss e 11, 79108 Freiburg, Germany}
\author{Mark O. Robbins}
\affiliation{Johns Hopkins University, Department of Physics and Astronomy,
  3400 North Charles Street, Baltimore, MD 21218, USA}

\begin{abstract}
At the molecular scale there are strong attractive interactions between surfaces, yet few macroscopic surfaces are
sticky.
Extensive simulations of contact by adhesive surfaces with roughness on nanometer to micrometer scales are used to determine how roughness reduces the area where atoms contact
and thus weakens adhesion.
The material properties, adhesive strength and roughness parameters are varied by orders of magnitude.
In all cases the area of atomic contact rises linearly with load,
and the prefactor rises linearly with adhesive strength for weak interactions.
Above a threshold adhesive strength,
the prefactor changes sign, the surfaces become sticky and a finite force is required to separate them.
A parameter-free analytic theory is presented that 
describes changes in these numerical results over up to five orders of magnitude in load.
It relates the threshold
strength to roughness and material properties, explaining why most macroscopic surfaces do not stick.
The numerical results are qualitatively and quantitatively inconsistent with classical theories based on the Greenwood-Williamson approach that neglect the range of adhesion and do not include asperity interactions. 
\end{abstract}


\maketitle

Surfaces are adhesive or ``sticky'' if breaking contact requires a finite force.
Few of the surfaces we encounter are sticky even though almost all are
pulled together by van der Waals interactions at atomic scales.~\cite{Israelachvili:3rd}
Gecko setae~\cite{Autumn:2000p681,Autumn:2002p12252} and engineered adhesives~\cite{Geim:2003p461} use this ubiquitous attraction
to achieve pull off forces per unit area that are orders of magnitude
larger than atmospheric pressure, and our world would come to a halt if these pressures operated on most macroscopic surfaces.

The discrepancy between atomic and macroscopic forces has
been dubbed the adhesion paradox.~\cite{kendallbook}
Experiments show that a key factor underlying this paradox is surface
roughness, which reduces the fraction of surface atoms that are close
enough to adhere.~\cite{Fuller:1975p327,kendallbook,perssonreview,Jacobs:2013}
Quantitative calculations of this reduction are extremely challenging
because of the complex topography of typical surfaces,
which have bumps on top of bumps on a wide range of scales.~\cite{archard70,Mandelbrot:Book1982}
In many cases they can be described as self-affine fractals from a lower
wavelength $\lambda_s$ of order nanometers to an upper wavelength $\lambda_L$
in the micrometer to millimeter range.~\cite{Mandelbrot:1984p721,perssonreview}
Here, we use an
efficient Green's function approach to calculate adhesive contact of surfaces with roughness
from subnanometer to micrometer scales. Numerical results for a wide range of surfaces,
adhesive interactions and material properties are presented and used to develop a simple,
parameter-free equation that predicts the effect of adhesion on contact.

The traditional Greenwood-Williamson (GW)~\cite{Greenwood:1966p300} approach for calculating contact of rough surfaces
 approximates their complex topography 
 by a set of spherical asperities of radius $R$ whose height distribution
is determined from self-affine scaling.
The long-range elastic interactions between different asperities are neglected.
This approach is analytically tractable and provided a simple explanation for the observation
that the area of contact between nonadhesive elastic surfaces is proportional to the normal force or load pushing them together.
Later generalizations\cite{Fuller:1975p327,Maugis:1996} considered the effect of adhesion between surfaces and found that the key parameter was the ratio of the root mean squared (rms)
height variation $h_{\text{rms}}$ to
the normal displacement $\delta_c$ of a single asperity due to adhesion.
If the work of adhesion gained per unit area of contact is $w$,
then $\delta_c^3 = (3/4)^3 R (\pi w/E^*)^2$
with contact modulus\cite{Johnson:Book1985} $E^*=E/(1-\nu^2)$ for an isotropic material with Young's modulus $E$
and Poisson ratio $\nu$.
GW based adhesion models\cite{Fuller:1975p327,Maugis:1996}
predict that
the force needed to separate surfaces drops rapidly as $h_\text{rms}/\delta_c$ increases
and is negligible for $ h_\text{rms}/ \delta_c >3$.

In the last decade, Persson has developed a scaling theory that includes an approximate
treatment of asperity interactions.\cite{Persson:2002p245502,Persson:2008p312001}
At the same time, large scale numerical calculations of contact between rough surfaces have become feasible.~\cite{Hyun:2004p026117,Campana:2007p38005,Campana:2008p354013,Putignano:2012p973}
Both approaches reveal limitations in the GW treatment of nonadhesive surfaces.
For example, the definition of $R$ is ambiguous,\cite{apology} the predicted range of linear scaling between area
and load is orders of magnitude too small \cite{Carbone:2008}, and predictions for the geomety of
individual contacts and the spatial correlations between them are qualitatively wrong.\cite{Campana:2008p354013,Persson:2008p312001}
As shown below, these geometrical features determine the effect of adhesion.

In recent work, Persson has extended his theory to include
adhesion in the limit where the range of surface separations
over which attractive interactions are significant, $\Delta r$, is zero.~\cite{Persson:2001,Persson:2002p245502,Peressadko:2005,Mulakaluri:2011}
He has applied this theory to specific cases and found a reduction in adhesion with increasing $h_\text{rms}$,\cite{} but this powerful approach
has not yet led to simple analytic predictions for general surfaces.

Here, we use an efficient Green's function approach to calculate adhesive
contact of surfaces with roughness from subnanometer to micrometer scales.
The numerical results are clearly inconsistent with expressions based on the GW approximation.
In particular, the relevant length scale describing the roughness is not $h_\text{rms}$ and the range of adhesive
interactions determines a characteristic adhesive
pressure $w/\Delta r$ that plays a critical role.
Numerical results for a wide range of surfaces, adhesive interactions
and material properties are presented and used to develop a simple,
parameter-free equation that predicts the effect of adhesion on contact.

\noindent
{\bf RESULTS }

\begin{figure*}[h]
  \begin{center}
    \includegraphics[width=16.5cm]{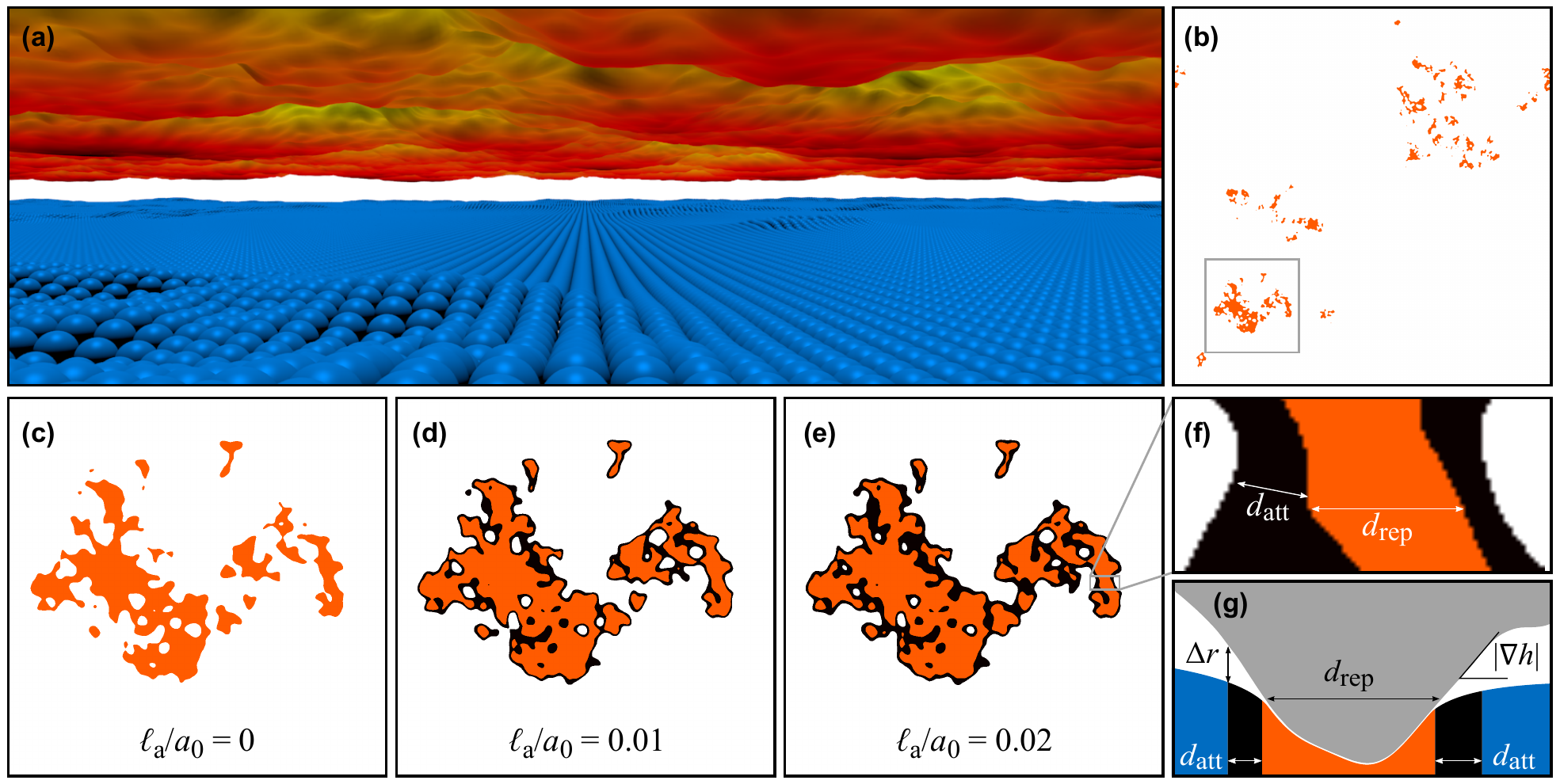}
    \caption{\label{geometry}
      {\bf Contact geometry.}
      {\bf (a)} 
The rigid top surface is a self-affine fractal with Hurst exponent $H=0.8$, root mean square slope $h_\text{rms}'=0.1$,
lower wavelength cutoff $\lambda_s=32 a_0$
and upper wavelength cutoff $\lambda_L=2048 a_0$.
The elastic substrate is initially flat with atoms spaced by $a_0$.
Substrate deformations produced by a typical adhesive contact are shown.
All height variations are magnified to show on the scale of the figure.
      {\bf (b-e)} Atoms that feel repulsive (orange) and attractive (black) forces at a fixed $A_\text{rep}\sim 0.02 A_0$. Nonadhesive results for the entire system are shown in (b) and expanded views of the region indicated by a square are
shown in (c-e) for the indicated $w/E^* a_0 = \ell_a/a_0$.
      {\bf (f)} Magnified view of an $80 a_0$ wide region of (e).
The mean diameter $d_\text{rep}$ is obtained by averaging the
distance across 
$A_\text{rep}$.
{\bf (g)} Vertical slice through a contact patch showing the rigid rough (gray)
 surface and the deformed elastic substrate.
 The root mean square slope, $h_\text{rms}'$, is the rms average of the local slope, $|\nabla h|$, as indicated on the right.
 The attractive length $d_\text{att}$
 is the distance from the repulsive patch edge at which the gap equals the
 interaction range $\Delta r$.
 The surface separation rises as a 3/2 power law for nonadhesive surfaces,
 leading to the 2/3 exponent in Eq.~\eqref{attrep}.
}
  \end{center}
\end{figure*}

Figure \ref{geometry}(a) shows the geometry of the simulations.
There is a rigid upper surface with self-affine roughness.
The change in height $h$ over a lateral distance $x$ increases
as $x^H$ where the Hurst exponent~\cite{Mandelbrot:Book1982}
 $H$ is between 0 and 1.
The elastic substrate
is the (100) surface of an fcc crystal
with atomic spacing $a_0$, and behaves like a continuous medium
in the limit of large $\lambda_s/a_0$.
We identify regions where atoms feel a repulsive force with
the contact area $A_\text{rep}$ (see methods).

\begin{figure}[h]
  \begin{center}
    \includegraphics[width=14cm]{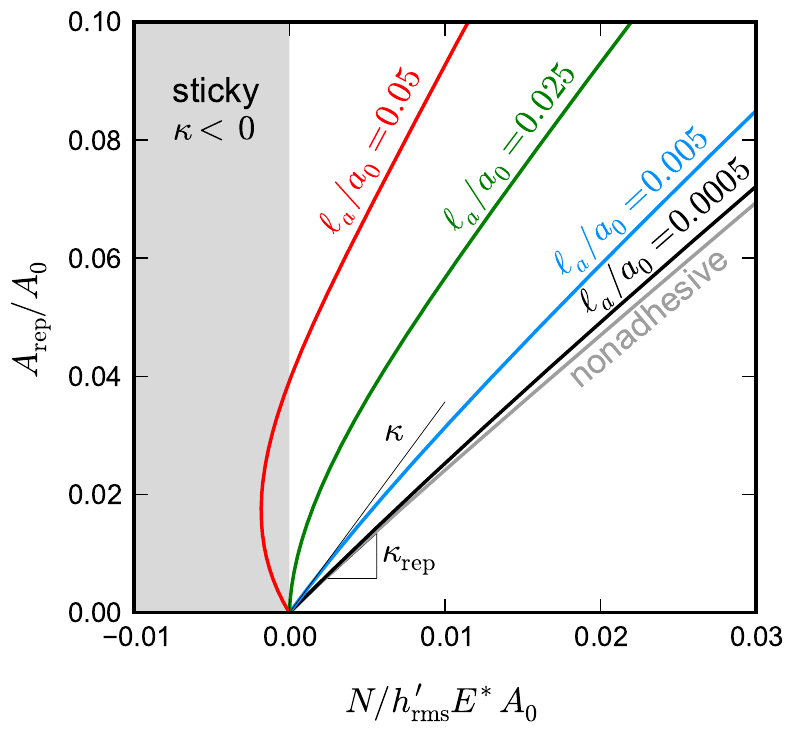}
    \caption{\label{area}
      {\bf Contact area as a function of normal load.}
      For nonadhesive contact, the contact area $A_\text{rep}$ rises
approximately linearly with load $N$ with dimensionless prefactor $\kappa_\text{rep}\approx 2$ (Eq. \ref{eqkappa}).
      As the strength of adhesion $\ell_a = w/E^*$ increases, the area rises
more rapidly with load.
      The initial ratio of area to load corresponds to a renormalized $\kappa$
that diverges at the onset of stickiness (green line).
      The red line shows a sticky case where $\kappa < 0$ and the pull-off force is nonzero.
      Results shown are for a surface with Hurst exponent $H=0.8$, root mean square slope $h_\text{rms}'=0.1$, and lower wavelength cutoff $\lambda_s=32 a_0$.
    }
  \end{center}
\end{figure}

Fig. \ref{geometry}(b) shows the complex spatial distribution of $A_\text{rep}$
for nonadhesive interactions.
Both GW and more recent approaches predict that $A_\text{rep}$ is much smaller than
the total area $A_0$ and rises linearly with the load $N$ pushing the surfaces
together.\cite{Greenwood:1966p300,Persson:2002p245502,Hyun:2004p026117,Campana:2007p38005,Persson:2008p312001,Putignano:2012p973}
By dimensional analysis, the surface geometry can only enter through
the dimensionless rms surface slope, $h_\text{rms}'=\sqrt{\langle |\nabla h|^2 \rangle}$ (see Fig.~\ref{geometry}g).
Steeper and stiffer surfaces are harder to bring into contact, so
that~\cite{Greenwood:1966p300,Persson:2002p245502,Hyun:2004p026117,Campana:2007p38005,Persson:2008p312001,Putignano:2012p973,Yastrebov:2012}
\begin{equation}
\frac{N}{A_\text{rep}} = \frac {h_\text{rms}' E^*}{\kappa_\text{rep}}  \equiv p_\text{rep} \ , 
  \label{eqkappa}
\end{equation}
where numerical solutions, such as the grey line in Fig.~\ref{area}, find
the dimensionless constant $\kappa_\text{rep}$ is close to 2 while GW and Persson give $\kappa_\text{rep} \approx 2.6$ and 1.6, respectively.
Note that the ratio of load to area represents the mean repulsive
pressure $p_\text{rep}$ in contacting regions, which depends
only on $h_\text{rms}'$ and $E^*$.

Figures \ref{geometry}(b)-(e) and \ref{area} show how adding adhesion
affects the distribution of contacting regions and the relation
between load and $A_\text{rep}$.
There is no need to separately consider the effect of $E^*$ and $w$ because
they always enter as a ratio with dimensions of length:
$\ell_a \equiv w/E^*$.
As discussed below, $\ell_a /a_0$ is typically much less than unity
and we use it to quantify the relative strength of adhesion.

Our first finding is that there is always a linear relation between the total load and the area in intimate repulsive contact at low $N$ (Fig. \ref{area}).
This can be described by Eq.~\eqref{eqkappa} with $\kappa_\text{rep}$ replaced by a renormalized constant $\kappa$. As the strength of adhesion increases, $\kappa$ and the ratio of $A_\text{rep}$ to load rise. Eventually the ratio diverges and the surfaces become sticky when $\kappa$ changes sign.
A negative $\kappa$ leads to an elastic instability that pulls surfaces into contact and a pulloff force equal to the magnitude of the most negative load (see Fig. \ref{area}) is needed to separate them.

\begin{figure}[h]
  \includegraphics[width=12cm]{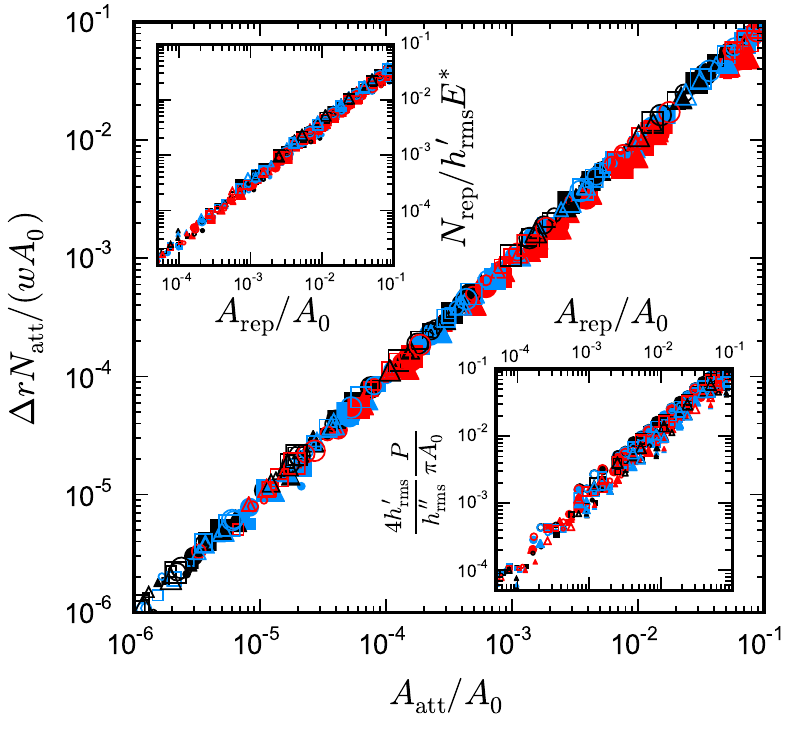}
  \begin{center}
    \caption{\label{att}
      {\bf Test of predictions for $N_\text{rep}$ and $N_\text{att}$.}
      The attractive load $N_\text{att}$ is equal to the attractive area $A_\text{att}$ times the mean adhesive pressure $w/\Delta r$.  Upper inset: The repulsive load $N_\text{rep}$ is equal to the repulsive area $A_\text{rep}$ times the mean repulsive pressure $p_\text{rep}\equiv h_\text{rms}' E^*/\kappa_\text{rep}$ where $\kappa_\text{rep}$ remains close to 2 even for the adhesive case with finite $\ell_a$.
      Lower inset: The length of the perimeter $P$ of the repulsive area $A_\text{rep}$ is proportional to the area itself.
      Plot is normalized to show that $d_\text{rep} \equiv \pi A_\text{rep}/P$ is nearly independent of $A_\text{rep}$ and $d_\text{rep}\approx 4h_\text{rms}'/h_\text{rms}''$.
Deviations by up to a factor of 2 from this expression for $d_\text{rep}$ are responsible for the spread in the figure. For a given system, changes in $d_\text{rep}$ with $A_\text{rep}$ are less than 25\% over 2-3 decades in $A_\text{rep}$.
All plots show multiple contact areas for each realization of a surface.
Results are shown for $H=0.3$ (triangles), 0.5 (squares) 
and 0.8 (circles)
and $\ell_a/a_0 = 0.0005$ (black), 0.005 (blue) and 0.05 (red).
Closed and open symbols are for $h_\text{rms}'=0.1$ and 0.3, respectively.
The symbol size increases as $\lambda_s/a_0$ increases from 4 to 64 
in powers of 2.
      }
  \end{center}
\end{figure}

A quantitative model for the changes in $\kappa$ can be derived
by analyzing how adhesion affects contact geometry.
Figures~\ref{geometry}(c-e) show contacting regions (orange) that
interact with repulsive forces and attracted regions (black) that are close 
enough to feel adhesive forces (Fig.~\ref{geometry}(f,g)).
The strength of adhesion is
varied at constant total repulsive area $A_\text{rep}$.
We find that the corresponding repulsive load $N_\text{rep}$ and
mean pressure $p_\text{rep}$ also remain constant
(see upper inset in Fig.~\ref{att})
and that there are only minor morphological changes in the shape of $A_\text{rep}$.
The main change is that the total area feeling an attractive
force, $A_\text{att}$,
spreads around the periphery of $A_\text{rep}$ as the range
of adhesive forces, $\Delta r$, increases
(Fig. \ref{geometry}(e,f)).
This type of behavior is assumed in the Derjaguin-Muller-Toporov (DMT) approximation for adhesion
which is typically valid for spherical asperities in the nanometer and micrometer range.\cite{Maugis:1992p243,Carpick:1999}
Different behavior might be observed if $\lambda_s$ was much larger
(Suppl. S-I).

The key observations needed to calculate $\kappa$ are that $p_\text{rep}$ remains constant, that
there is a constant mean adhesive pressure $w/\Delta r$ in the attractive region (see Fig.~\ref{att})
and that the ratio of repulsive and attractive areas is independent of load
at low loads.
The first two observations allow us to write the total load as $N=p_\text{rep} A_\text{rep}-(w/\Delta r) A_\text{att}$.
 From Eq.~\eqref{eqkappa} we immediately find
\begin{equation}
1/\kappa = 1/\kappa_\text{rep}-1/\kappa_\text{att}
\end{equation} with
$\kappa_\text{rep} \approx 2$ and
\begin{equation}
  {\kappa_\text{att}}
  =
  h_\text{rms}'\frac{\Delta r}{\ell_a} \frac{A_\text{rep}}{A_\text{att}} \ \ .
\label{eqrenkappa}
\end{equation}
The remaining unknown is the ratio of repulsive to attractive area.

If $A_\text{att}\ll A_\text{rep}$,
it can be approximated by
$A_\text{att} \approx P d_\text{att}$ where
$P$ is the length of the perimeter of $A_\text{rep}$ and
$d_\text{att}$ the average
lateral distance from the perimeter where the surface separation reaches
the interaction range $\Delta r$ (Fig. \ref{geometry}(g)).
For a general area,
$A_\text{rep} = P d_\text{rep}/\pi$
where $d_\text{rep}$ is the mean diameter
(see Fig. \ref{geometry}(f) and Suppl. S-II).
For ordinary two dimensional objects like a circle, the perimeter
and diameter
are proportional and increase as the square root
of the area.
This behavior is assumed in conventional theories of contact between rough surfaces that ignore
long-range elastic interactions between individual contacting asperities, such as the
previously discussed GW~\cite{Greenwood:1966p300}
and related adhesive~\cite{Fuller:1975p327,Maugis:1996,Chow:2001p4592} models.
Including elastic interactions leads to qualitative changes in contact geometry~\cite{Hyun:2004p026117,Persson:2008p312001,Campana:2008p354013}.
The contact area becomes a fractal with the same fractal dimension as the perimeter~\cite{Hyun:2004p026117}
(a true ``monster''~\cite{Mandelbrot:Book1982}).
We find that $d_\text{rep}$ is then nearly independent of
contact area, load and adhesive strength and present an analytic expression
for it below.

We calculate $d_\text{att}$ using continuum theory
for nonadhesive contact of locally smooth surfaces.
If $x$ is the lateral distance from the edge of a contact, then for small $x$ the separation between surfaces always\cite{Johnson:Book1985} rises as $x^{3/2}$.
We use the standard prefactor for a cylinder
with contact radius $d_\text{rep}/2$
and equate $d_\text{att}$ to the lateral distance where the separation
equals $\Delta r$ (Suppl. S-I).
Combined with our expression for $P$, this gives
the constant ratio between attractive and repulsive areas
\begin{equation}
  \frac{A_\text{rep}}{A_\text{att}}
  =
\frac{d_\text{rep}}{\pi d_\text{att}}
=
  \left[\frac{16}{9\pi}\right]^{1/3}
  \left[\frac{h_\text{rms}' d_\text{rep}}{\pi\Delta r}\right]^{2/3} \ .
  \label{attrep}
\end{equation}
Inserting this result in Eq.~\eqref{eqrenkappa}, gives the prediction for $\kappa_\text{att}$.

\begin{figure}[h]
  \begin{center}
    \includegraphics[width=14cm]{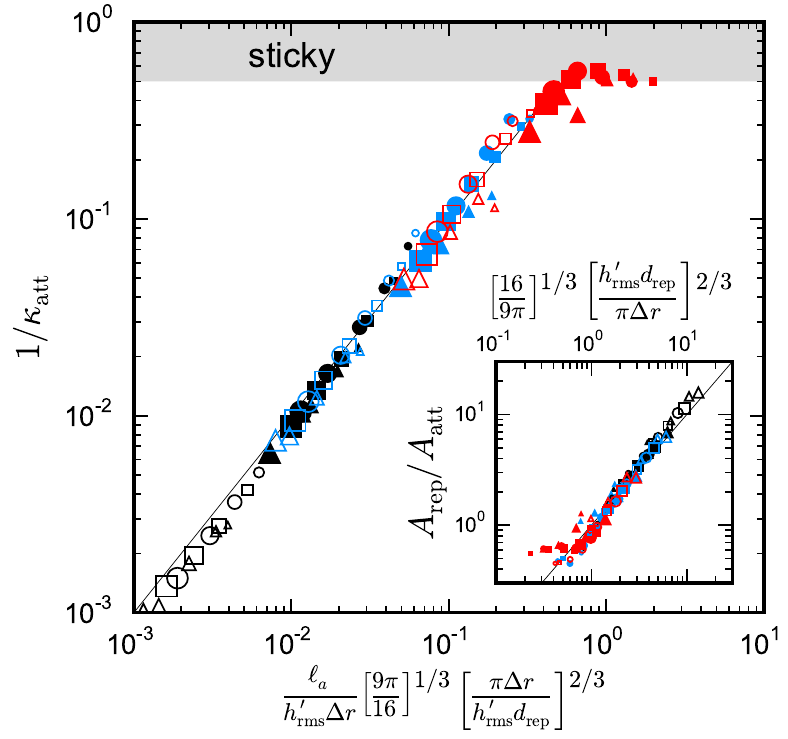}
    \caption{\label{kappa}
      {\bf Test of predictions for $\kappa_\text{att}$ and $A_\text{att}$.}
      The attractive $1/\kappa_\text{att}$ is plotted against the prediction in Eq.~\eqref{eqrenkappa}. The solid line has unit slope.
Results are shown for Hurst exponents $H=0.3$ (triangles), 0.5 (squares) 
and 0.8 (circles)
and adhesive strengths $\ell_a/a_0 = 0.0005$ (black), 0.005 (blue) and 0.05 (red).
Closed and open symbols are for root mean square slopes $h_\text{rms}'=0.1$ and 0.3, respectively.
The symbol size increases as the lower wavelength cutoff $\lambda_s/a_0$ increases from 4 to 64 in powers of 2.
Adhesion is observed if and only if
$1/\kappa_\text{att} > 1/\kappa_\text{rep}$.
The shaded region shows the prediction for adhesion using the value
$\kappa_\text{rep}=2$ found in the continuum limit (Suppl. S-2).
Inset: Plot of the ratio
of repulsive to attractive area $A_\text{rep}/A_\text{att}$
against the prediction of
Eq.~\eqref{attrep}.
    }
  \end{center}
\end{figure}

As shown in Fig.~\ref{kappa}, our simple analytic expressions provide
a quantitatively accurate description of $A_\text{rep}/A_\text{att}$ and
$\kappa_\text{rep}/\kappa_\text{att}$
for a wide range of surface geometries.
Deviations are only larger than the numerical uncertainty
when the attractive area has grown too large to be approximated
as a thin rim around the repulsive region (i.e. when $A_\text{att} > A_\text{rep}$), which is well into the sticky regime.
The continuum expression for $d_\text{att}$ also fails
in this limit ($d_\text{att} > d_\text{rep}/2$).

Eqs.~\eqref{eqrenkappa} and \eqref{attrep} provide a simple and quantitative explanation for the changes in Fig.~\ref{area}.
As the adhesion energy (and therefore $\ell_a$) increases,
there is a proportional increase in $1/\kappa_\text{att}$.
At first adhesion merely produces a small change
in the ratio of area to load.
The sign of the ratio changes when
$1/\kappa_\text{att}$ becomes bigger than $1/\kappa_\text{rep}$ 
and the surface becomes sticky.  

The length $d_\text{rep}$ is always of order $\lambda_s$ and has a simple relation to statistical properties
of the undeformed surface.
As above and in Suppl. S-I, we approximate the contacting part of asperities by a cylinder with radius $R$, which is calculated from the rms curvature of the rough
surface $1/R=h_\text{rms}^{\prime\prime}/2=\sqrt{\langle (\nabla^2 h)^2\rangle}/2$.
If the contact has diameter $d_\text{rep}$ and slope $h_\text{rms}^\prime$ at the contact edge, then $d_\text{rep}=4 h_\text{rms}' / h_\text{rms}^{\prime \prime}$.
Following the same reasoning, the
length in the numerator of Eq.~\eqref{attrep} is proportional
to the height change $\delta h$ from the contact
edge to center:
\begin{equation}
 \delta h = \left[ h_\text{rms}^{\prime} \right]^2/h_\text{rms}^{\prime\prime} = h_\text{rms}' d_\text{rep}/4  
\end{equation}
The values of $h_\text{rms}^\prime$ and $h_\text{rms}^{\prime\prime}$
can readily be evaluated from the statistical properties of the
rough surface in real or reciprocal space.
The lower inset of Fig.~\ref{att} shows that $P$ and $A_\text{rep}$ are proportional and that the proportionality constant is always within a factor of two of $4 h_\text{rms}' / h_\text{rms}^{\prime \prime}$.

The contact area is not directly accessible to experiment, but changes with load in the mean separation between surfaces $u$ can be measured with sufficiently stiff mechanical devices.~\cite{Oliver:1992,Carpick:1997}
The normal contact stiffness defined as $k_N =d N /d u$ is typically found to
rise linearly with load for nonadhesive surfaces.~\cite{Akarapu:2011,Campana:2011}
In the regime where surfaces are not sticky we find that the relation between surface separation and $N_\text{rep}$ is nearly unchanged, just as the relation between $N_\text{rep}$ and $A_\text{rep}$ is nearly the same (Fig. 3 upper inset).
Since adhesion reduces the total load $N$ by a factor of $\kappa_\text{rep}/\kappa$, the normal stiffness is reduced by the same factor.
This is a small correction unless the surfaces are close to becoming sticky,
and nonadhesive predictions are likely to be within experimental uncertainties.

\noindent
{\bf DISCUSSION}

Surfaces are sticky when the total adhesive force, which is adhesive pressure
times attractive area, exceeds the total repulsive force $p_\text{rep}A_\text{rep}$.
This corresponds to $1/\kappa_\text{att}>1/\kappa_\text{rep}$, and our
numerical results show stickiness if and only if this condition is met.
It can be recast as a condition on the ratios of pressures and areas,
$(w/\Delta r)/p_\text{rep} > A_\text{rep}/A_\text{att}$, or using
our analytic expressions:
\begin{equation}
  \frac{h_\text{rms}' \Delta r}{\kappa_\text{rep} \ell_a} \left[\frac{\delta h}{\Delta r}\right]^{2/3}
   < \pi \left[\frac{3}{16}\right]^{2/3} \approx 1
  \label{adcond}
\end{equation}
where the first factor on the left reflects the pressure ratio 
and the second comes from the area ratio.
As noted above and in the supplementary material
the effective range of the potential is typically less than but
of order of the atomic bond-distance $a_0$.
Any height change $\delta h$ is possible in continuum theory,
but there is a natural lower bound of order $a_0$ in atomic systems.
For example, roughness on crystalline surfaces occurs in the form of terraces with height $\sim a_0$.
Inserting this bound in Eq.~\eqref{attrep}
one finds
 a necessary but not sufficient criterion for adhesion:
$\ell_a/a_0 \gtrsim 0.5 h_\text{rms}'$.

Note that the above prediction for the onset of adhesion is
qualitatively different than previous models for rough surface
adhesion, which do not include two of the key lengths in Eqs.~\eqref{eqrenkappa}--\eqref{adcond}.
The characteristic width of contacting regions $d_\text{rep}$
reflects their fractal nature and has not been identified before.
Continuum theories have considered the limiting cases of $\Delta r$
equal to zero\cite{Fuller:1975p327,Chow:2001p4592,Persson:2002p245502}
or infinity\cite{Maugis:1996} and concluded $\Delta r$ had little effect,\cite{Maugis:1996}
while we find more adhesion at small $\Delta r$ because the adhesive pressure
is increased.
Finally, our relations only include quantities that are determined by
short wavelength roughness -- the rms surface slope and curvature.
The rms roughness is the key surface property in past GW theories,
and rises with the upper wavelength of roughness as $\lambda_L^H$.
The numerical results with different symbol size in Fig.~\ref{kappa} 
have values of $\lambda_L^H$ that vary by more than an order of magnitude
but collapse onto the universal curve predicted by Eqs.~\eqref{eqrenkappa}
and \eqref{attrep}.
Supplemental section S-III presents plots that show qualitative discrepancies between these data and traditional GW theories.

To determine the implications of Eq.~\eqref{adcond},
we first consider the extreme case where
$w$ is the adhesive energy for joining crystals of the same material.
Then for atomistic solids the same interactions determine both
$E^*$ and $w$.
The value of $\ell_a$ is of order of the relative displacement needed
to change the elastic energy by the binding energy, and $\ell_a/a_0 \ll 1$.
For example, diamond has $E^* \approx 10^{12}$Pa and
$w \approx 10\,{\rm J/m^2}$, yielding
$\ell_a/a_0 \approx 0.06$ with $a_0$ the carbon bond length.
The simple Lennard-Jones potential has $\ell_a/a_0 \approx 0.05$.
For these typical values of $\ell_a$,
adhesion should occur for $h_\text{rms}'$ of order 0.1 and below.
The stickiest cases considered in Fig.~\ref{kappa} (closed red symbols)
are indeed for the case
$\ell_a/a_0=0.05$, $h_\text{rms}'=0.1$, and small $d_\text{rep}$.
Increasing $h_\text{rms}'$ to 0.3 suppresses adhesion.

Exposing surfaces to the environment typically reduces the adhesive forces
to van der Waals interactions
with $w \sim 50$mJ/m$^2$.
The value of $\ell_a$ and the root mean square slope needed to eliminate adhesion
are reduced by two to three orders of magnitude.
Only exceptionally smooth surfaces like atomically flat
mica~\cite{Israelachvili:3rd}
and the silicon surfaces used in wafer bonding~\cite{Tong:1999p1409}
have slopes low enough to stick ($\lesssim 10^{-3}$).
For most surfaces $\kappa \approx \kappa_\text{rep}$.
This provides an explanation for the success of models
for friction that ignore
van der Waals adhesion.~\cite{Bowden:Book1950,Greenwood:1966p300,Johnson:Book1985,Carpick:1999}

Most of the sticky surfaces we are familiar with break the connection
between $w$ and $E^*$ to increase $\ell_a/a_0$.
Geckos~\cite{Autumn:2000p681,Autumn:2002p12252} and recently manufactured mimics~\cite{Geim:2003p461} break
the solid up into a hierarchical
series of separate rods with pads at the ends.
This allows adjacent pads to contact the surface at different heights without a large elastic energy, leading to a small effective $E^*$ even though the components are stiff.
Tape, rubber, and elastomers adhere via van der Waals interactions, 
but have small elastic moduli associated with the entropy
required to stretch polymer segments between chemical crosslinks.
Eq.~\eqref{adcond} implies that surfaces with $w=50$mJ/m$^2$, $h_\text{rms}' \sim 1$, $\Delta r \sim 0.5\,{\rm nm}$ and $d_\text{rep} \sim 10\,{\rm nm}$ will be sticky if $E^* \lesssim 10\,{\rm MPa}$, which is
common for soft rubbers and elastomers while paper is much stiffer ($>$1GPa).
Tapes are normally designed to have moduli below 0.1MPa, which is known
as the Dahlquist criterion \cite{Creton:1996p545}.
Taking $h_\text{rms}' \sim 1$ and $\Delta r \sim 0.5\,{\rm nm}$, one finds
adhesion for $d_\text{rep} \lesssim 100\,{\rm \mu m}$.
Adhesives of this type can stick even to structured surfaces with macroscopic
grooves.
Once an adhesive bond is formed, 
the viscoelastic properties of the adhesive that can be used to greatly increase the force needed to break the adhesive bond~\cite{kendallbook}.

In summary, we have presented numerical simulations of adhesive contact between rough surfaces for a wide range of adhesion strength, surface geometries and material properties.
In all cases the area in intimate
repulsive contact rises linearly with the applied load at low loads.
The ratio of area to load increases with adhesion and changes sign
when surfaces begin to stick.
This transition only occurs in the limits of smooth surfaces, high surface
energy and low stiffness.
The results are qualitatively inconsistent with traditional GW theories, but
in quantitative agreement with a simple parameter-free theory
based on observed changes in contact geometry.
This theory makes specific predictions for experimental systems and may aid
in the design of adhesives, and in engineering surface roughness to
enhance or eliminate adhesion.
It also provides a simple explanation for our everyday experience with
macroscopic adhesion.
For most materials the internal cohesive interactions that determine elastic
stiffness are stronger than adhesive
interactions and surfaces will only stick when they are extremely smooth.
Tape, geckos and other adhesives stick because the effect of internal bonds
is diminished to make them anomalously compliant.
\newpage

\emph{Methods ---}
Calculations were performed for a rigid rough surface contacting a flat
elastic substrate.
In continuum theory this is equivalent to contact between two rough elastic
surfaces and the mapping remains approximately valid at atomic scales.~\cite{Luan:2005p929}
Self-affine rough surfaces with the desired Hurst exponent $H$, $h_\text{rms}'$, $\lambda_s$ and $\lambda_L$
are generated using a Fourier-filtering algorithm described
previously.~\cite{Ramisetti:2011p215004}
Fourier components for each wavevector $\vec{q}$ have a random phase and a normally distributed amplitude that depends on the magnitude $q$.
The amplitude is zero for $q > 2\pi/\lambda_s$, proportional to $q^{-1-H}$ for
$2\pi/\lambda_s>q>2\pi/\lambda_L$, and rolls over to a constant for $q < 2\pi/\lambda_L$.
Periodic boundary conditions with period $L$ are applied in the plane of the surface to prevent edge effects.
The elastic response is determined using a Fourier-transform technique~\cite{Pastewka:2012p075459,Campana:2006p075420} with a linearised surface Green's function
corresponding to Poisson ratio $\nu=1/2$.
Results are shown for period $L=2\lambda_L =4096a_0$ with rigid boundary conditions applied at depth $L$ below the surface.
Systematic studies were performed with $L$ and $\lambda_L$ from 512$a_0$ to 8192$a_0$ to ensure that finite-size effects are small.

Atoms on the elastic substrate interact with the rigid rough surface through
a potential that only depends on the height difference $z$.
We use a $9$-$3$ Lennard-Jones potential that represents the integral over a half space of the usual $12$-$6$ Lennard-Jones potential between atoms.
The potential is truncated smoothly using a cubic spline from
the potential minimum at $z=a_0$ to the cutoff at $a_0+\Delta r$.
The potential and its first two derivatives are continuous and vanish
at the cutoff.
In our calculations we fix the stiffness $k$ of the potential at a value
that is consistent with the stiffness of interactions within the substrate:
$k=E^* a_0/2$.
Consistent results were obtained with other potentials, including
an untruncated $12$-$6$ Lennard-Jones potential.

As is common for atomic-scale calculations~\cite{Mo:2009p1116,Knippenberg:2008p235409,Luan:2005p929}, the contact area $A_\text{rep}$ is defined as the area covered by atoms that feel a repulsive force ($z<a_0$). Similarly, the attractive area $A_\text{att}$ is the area covered by atoms that feel an attractive force ($a_0<z<a_0+\Delta r$).
We only show results for $A_\text{rep}/\lambda_s^2 \gtrsim 10$, so that there is a statistical number of contacting asperities.\cite{Hyun:2004p026117}
Numerical values of $\kappa$, $\kappa_\text{rep}$ and $\kappa_\text{att}$ are computed at $1\%$ contact area from the ratios of load and area.

\newpage

\emph{Acknowledgements ---}
This work was supported by the Air Force Office of Scientific Research (grant FA9550-0910232), the U.S. National Science Foundation (grant OCI-108849, DMR-1006805, CMMI-0923018), the Simons Foundation (M.O.R.)
and the European Commission (Marie-Curie IOF 272619 for L.P.).
Computations were carried out at Johns Hopkins University and the J\"ulich Supercomputing Center.

\newpage

\clearpage

\setcounter{figure}{0}
\setcounter{equation}{0}

\renewcommand{\thesection}{S-\Roman{section}}
\renewcommand{\thefigure}{S-\arabic{figure}}
\renewcommand{\thetable}{S-\arabic{table}}
\renewcommand{\theequation}{S-\arabic{equation}}

\begin{center}
\huge\bf{ Supplementary Material for \\
 ``When are rough surfaces sticky?'' }

\vspace{1cm}

\large Lars Pastewka$^{1,2}$ and Mark O. Robbins$^1$
\end{center}

$^1$ Johns Hopkins University, Dept. Physics and Astronomy,
  Baltimore, MD 21218, USA

$^2$ Fraunhofer IWM, MikroTribologie Centrum $\mu$TC, 79108 Freiburg, Germany

\section{Surface separation, $d_\text{att}$, and the effective range $\Delta r$ for arbitrary interaction potentials}
\label{sep}

As discussed in the main text, we use standard results from
continuum theory for nonadhesive contact between 
smooth surfaces.
If $x$ is the lateral distance from the edge of the contact, then for small $x$ the separation
$\Delta(x)$ between surfaces always rises as $x^{3/2}$
outside the contact.\cite{Johnson:Book1985_2}
The prefactor rises with the surface slope at the edge of the contact
which we take to be $h_\text{rms}'$.
For simple geometries like spheres, cones or cylinders, the only length scale that enters is the radius of the contact area.
Since the contacting region in our numerical simulations has a constant average diameter $d_\text{rep}$, we use the standard prefactor for a cylinder:~\cite{Baney:1997p393,Yang:2009p1976}
\begin{equation}
   2 \frac{\Delta(x)}{d_\text{rep}} = \frac{\sqrt{8}}{3} h_{\rm rms}' \left(\frac{2x}{d_\text{rep}}\right)^{3/2}
  \label{gap}
\end{equation}
Our numerical data for the average surface separation at a given
distance from the perimeter are consistent with this relation without adjustable parameters.
To find $d_\text{att}$
we just equate $\Delta(d_\text{att})$ to $\Delta r$ in Eq.~\eqref{gap},
yielding
\begin{equation}
  d_\text{att}/d_\text{rep}
  =
  \left( \frac{3}{4} \frac{\Delta r}{ h_{\rm rms}' d_\text{rep}} \right)^{2/3}.
\label{datt}
\end{equation}

An effective range of interaction $\Delta r$ can be defined for arbitrary forms
of the interaction potential using Eq.~\eqref{gap}.
We define $p(\Delta)$ as the attractive pressure between surfaces separated by $\Delta > 0$.
Then $w = \int\limits_0^\infty \mathrm{d}\Delta\, p(\Delta)$ where $w$ is the work of adhesion.
As in the calculation of $A_\text{att}$ in the main document, we assume that the perimeter changes direction slowly enough that we can
write the total load as the perimeter times a contribution per unit length.
With $\Delta(x)$ being the separation at distance $x$ from the contact edge, this yields:
\begin{equation}
  N_\text{att}=P \int\limits_0^\infty \mathrm{d}x\, p(\Delta(x)) \ \ .
\label{Natt}
\end{equation}
This is equivalent to the expressions in the main text for truncated potentials with
\begin{equation}
  d_\text{att} / \Delta r = \int\limits_0^\infty \mathrm{d}x\, p(\Delta(x))/w \ \ .
  \label{datt2}
\end{equation}
From Eqs.~\eqref{gap}, \eqref{datt} and \eqref{datt2} one finds
an expression for $\Delta r$
\begin{equation}
  \left( \Delta r \right)^{-1/3}=\left.\frac{2}{3} \int\limits_0^\infty \mathrm{d} \Delta\, \Delta^{-1/3} p(\Delta) \middle/
  \int\limits_0^{\infty} \mathrm{d} \Delta\, p(\Delta) \ \ .\right.
\label{deltar}
\end{equation}
With this value of $\Delta r$, all of the relations in the main text for truncated potentials carry over to an arbitrary potential.
Note that the integrals are well defined because $p$ represents
the total force and goes linearly to
zero at the equilibrium surface separation $\Delta=0$.

In general we find that $\Delta r$ from Eq.~\eqref{deltar} is comparable to or smaller than the
atomic spacing $a_0$ that minimizes the energy.
For example, if the 9-3 Lennard-Jones potential is used to infinite distances,
one finds $\Delta r = 1.15 a_0$.
For the 12-6 Lennard-Jones potential, $\Delta r = 0.62 a_0$.
Note that these ranges are referenced to the potential minimum at $a_0$ and that
most of the binding energy comes from these short scales ($>85$\%).

The spline potential used in the calculations reported in the main text has
\begin{equation}
  p(z)
  =
  \frac{k}{a_0^{2}}
  \left\{
  \begin{array}{ll}
    \frac{a_0}{6}
    \left[
      \left(\frac{a_0}{z}\right)^{10}
      -
      \left(\frac{a_0}{z}\right)^4
    \right]
    &
    \quad\text{if}\quad z \leq a_0
    \\
    -(z-a_0)
    +
    \frac{2}{\Delta r} (z-a_0)^2
    -
    \frac{1}{\Delta r^2} (z-a_0)^3
    &
    \quad\text{if}\quad a_0 < z < a_0+\Delta r
    \\
    0
    &
    \quad\text{if}\quad z > a_0+\Delta r
  \end{array}
  \right.
  \label{potential}
\end{equation}
where $a_0$ is the minimum of the potential
and $k=E^* a_0/2$ is the stiffness at $z=a_0$.
The adhesion energy was varied by changing $\Delta r$ at fixed $k$.
The range given by Eq.~\eqref{deltar} for this potential is essentially
the same as the actual range (within 6\%).
For a general truncated potential, quantitative agreement with results for the relation between area and load are better when Eq.~\eqref{deltar} is used.

In calculating the load we have assumed that the variation of the surface
separation $\Delta$ with $x$ is not affected by adhesion.
A similar approximation is made in the Derjaguin-Muller-Topov (DMT) theory for contact of a spherical asperity of radius $R$.~\cite{Derjaguin:1975p314}
Maugis has found that the DMT approximation is accurate for spheres
when a dimensionless ratio $\lambda_\text{Maugis}$ is small:~\cite{Maugis:1992p243_2}
\begin{equation}
\lambda_\text{Maugis} \equiv \left[ \frac{9 R \ell_a^2}{2\pi (\Delta r)^3} \right]^{1/3} \ \ .
\end{equation}
As above and in the main text, we determine $R = d_\text{rep}/2h_\text{rms}^\prime$
by assuming a locally cylindrical geometry with contact diameter $d_\text{rep}$ and slope $h_\text{rms}^\prime$ at the edge of the contact.
We find that the approximations used in the main text are accurate
even when $\lambda_\text{Maugis}$ exceeds unity.
Using typical values of $\ell_a \sim 5 \cdot 10^{-4}$ for atomistic solids exposed to the environment,
$h_\text{rms}^\prime=0.1$ and $\Delta r/a_0 = 1$,
this corresponds to $d_\text{rep}$ of close to a millimeter.
For elastomers and other compliant systems, surface deformation
becomes important at smaller scales, but the surfaces are usually
sticky and other corrections are also required.
For example, $A_\text{att}$ is typically not small compared to $A_\text{rep}$.
Previous analytic studies of rough surface adhesion have considered the
Johnson-Kendall-Roberts (JKR) limit
where the area outside the contact does not contribute significantly to the adhesion.~\cite{Johnson:1971p301}
This corresponds to $\lambda_\text{Maugis} > 5$ where the contact radius becomes of macroscopic dimensions.

Finally, many of the above relations assume that the surface slope is relatively small so that the total and projected areas are nearly equal and the potential interaction depends only on the vertical separation.
The strains induced by contact are of order the surface slope and plasticity
also becomes important for very steep surfaces.

\section{Determining contact patch geometries}

For continuous curves,
the perimeter and area are related through $d_\text{rep}$, the mean length of contiguous segments in horizontal
(or vertical) slices through $A_\text{rep}$ (see Fig.~1(f) in main text).
Suppose the slices are made at a spacing $dz$ that can be made arbitrarily small.
The total area can be approximated by the sum over all contiguous segments
of the segment length times $dz$.
If the total number of segments is $N_\text{tot}$, then
$A_\text{rep}= N_\text{tot} d_\text{rep} dz$.
Each end of a segment contributes $dz$ to the
projection of the perimeter perpendicular to the slice.
The projected perimeter length is then $2N_\text{tot}dz$.
The total perimeter length is $\pi N_\text{tot} dz$ if one assumes that all orientations
are sampled equally -- as for a circle.
This gives the relation $A_\text{rep}=P d_\text{rep}/\pi$ cited in the text.

\begin{figure}
  \begin{center}
    \includegraphics[width=12cm]{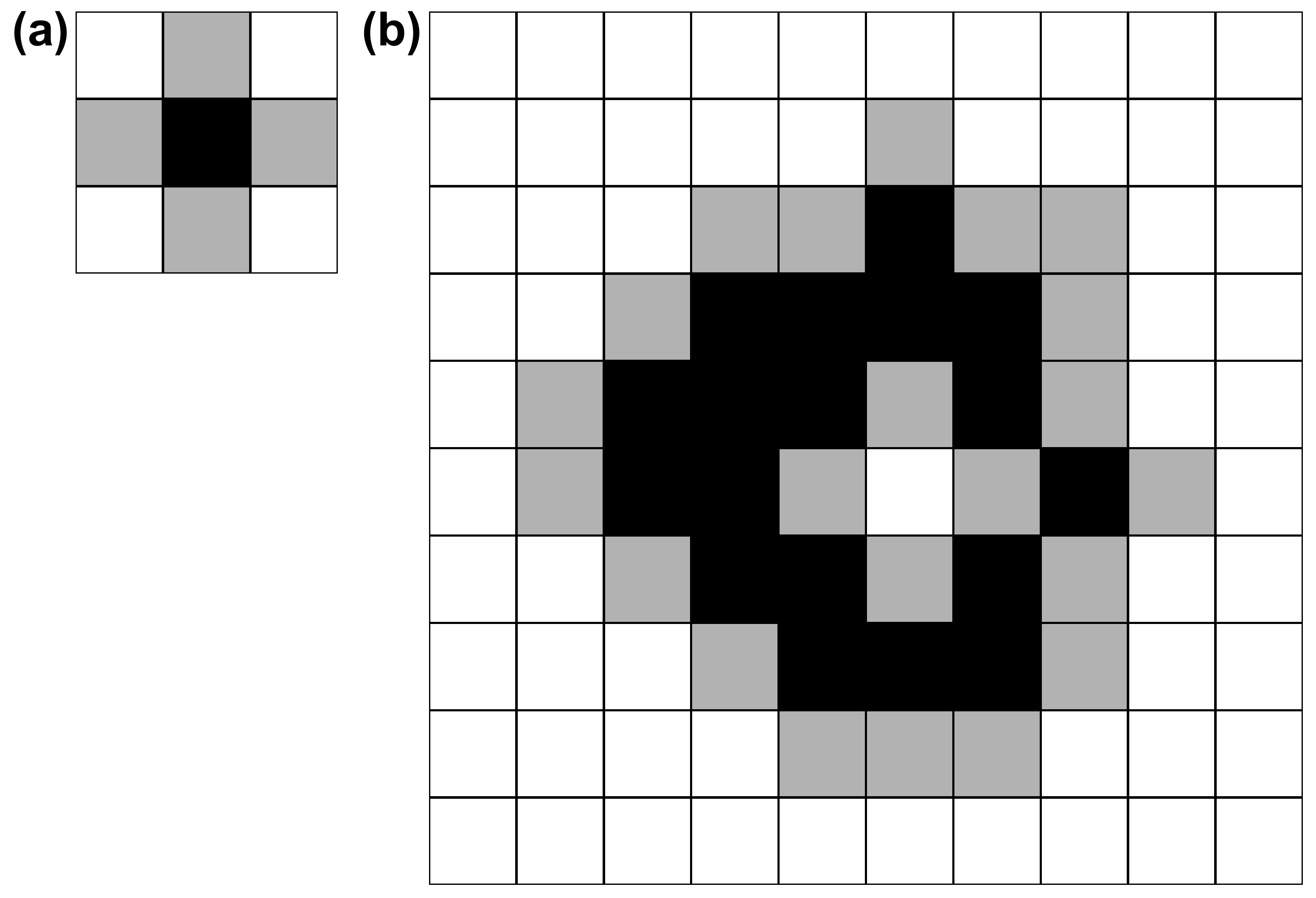}
    \caption{\label{neighbors}
      {\bf Identifying the contact patch perimeter.}
      {\bf (a)} Grid cells are neighbors when they share an edge.
      Gray squares are neighbors of the black square in the middle.
      {\bf (b)} A patch showing perimeter cells with less than four neighbors in gray and interior cells with four neighbors in black.
      }
  \end{center}
\end{figure}

Figure~\ref{neighbors} illustrates how area and perimeter are defined on the discrete geometry
used in our simulations.
Atoms on the substrate surface form a square grid with spacing $a_0$.
The surface is divided into a grid of square cells centered on each atom.
A grid cell contributes $a_0^2$ to $A_\text{rep}$ if the corresponding atom feels a repulsive force and
to $A_\text{att}$ if the force is attractive.

Grid cells are defined to be neighbors if they share an edge (Fig.~\ref{neighbors}a).
The corresponding atoms are then nearest neighbors.
The repulsive area is divided into connected patches like that shown in Fig.~\ref{neighbors}b.
Grid cells (atoms) that belong to a patch but have less than four neighbors in the patch are part of the perimeter.
The perimeter length $P$ is calculated as: $P= \beta a_0 N_P$ where
$N_P$ is the number of perimeter cells and $\beta$ corrects for the discreteness
of the lattice.
Consider a line of length $L$ at an angle $\theta$ to the horizontal axis.
The perimeter cells will form a stepped approximation to this line.
It is easy to show that the number of perimeter cells is equal to $L/(a_0 \cos \theta)$ for $\theta$ between $-\pi/4$ and $\pi/4$.
Counting these cells and multiplying by $a_0$ underestimates $P$ by a factor of
$1/\cos \theta$.
We find $\beta=4 \sinh^{-1} (1) /\pi \approx 1.1222$ by assuming isotropy and averaging over angles.

\begin{figure}
  \begin{center}
    \includegraphics[width=12cm]{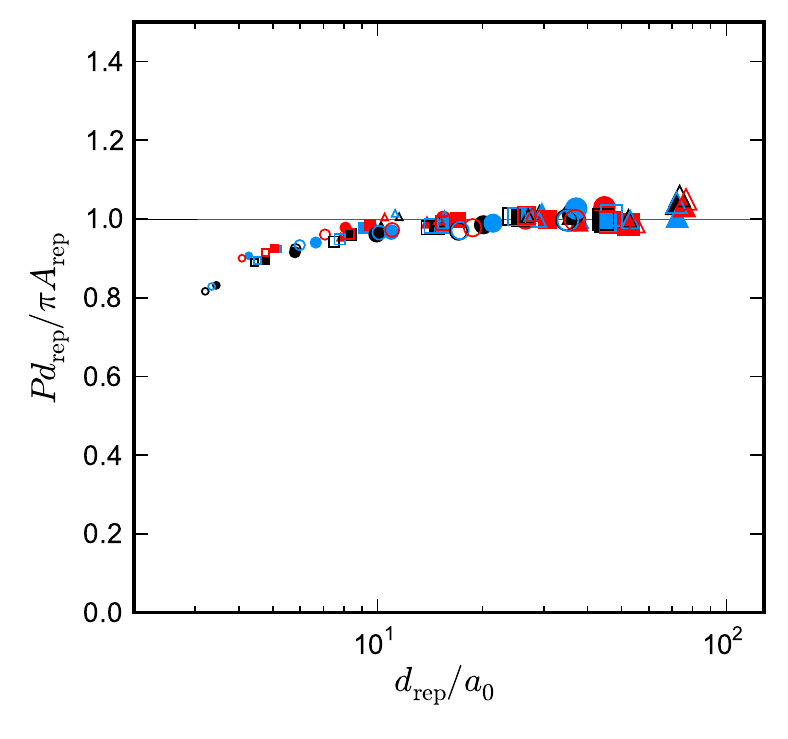}
    \caption{\label{perimeter}
    {\bf Test of relationship between perimeter $P$, mean diameter $d_\text{rep}$ and area $A_\text{rep}$.}
    The ratio of $P d_\text{rep}/\pi$ to
repulsive area
as a function of the mean width
of contacting regions $d_\text{rep}$.
Results are shown for $H=0.3$ (triangles), 0.5 (squares) 
and 0.8 (circles)
and $\ell_a/a_0 = 0.0005$ (black), 0.005 (blue) and 0.05 (red).
Closed and open symbols are for $h_{\rm rms}'=0.1$ and 0.3, respectively.
The symbol size increases as $\lambda_s/a_0$ increases from 4 to 64 
in powers of 2.
      }
  \end{center}
\end{figure}

Figure~\ref{perimeter} tests the above relations by plotting the predicted ratio
of perimeter to area as a function of $d_\text{rep}$ for the full range of $H$, $\lambda_s$
and $w$ discussed in the main text.
Agreement is excellent in the continuum limit.
For $d_\text{rep}/a_0 > 10$ results are within the numerical uncertainty.
The deviation at the greatest $d_\text{rep}$ is reduced if $\lambda_L/d_\text{rep}$ is
increased to remove finite-size effects.
As $d_\text{rep}/a_0$ decreases below 10, there is an accelerating drop in the plotted
ratio.
The above relations assumed lines were straight, 
and curvature can not be ignored when the radius of curvature of
the perimeter, $\sim d_\text{rep}/2$, is comparable to the grid spacing, $a_0$.

\section{Comparison to Theories Based on the Greenwood-Williamson Approximation}

As noted in the main text, traditional theories for the effect of adhesion start
from the Greenwood-Williamson approximation.
Fuller and Tabor\cite{Fuller:1975p327_2} found the pulloff force needed to separate the surfaces
in the JKR limit of short range potentials ($\Delta r \rightarrow 0$) and
Maugis\cite{Maugis:1996_2} found similar results for the opposite DMT limit of long range
potentials ($\Delta r \rightarrow \infty$).
In both cases, the pulloff force is a function of 
$h_\text{rms}/\delta_c$,
where $\delta_c$ is the normal displacement of a single asperity
due to adhesion $\delta_c^3=(3/4)^3R(\pi w/E^*)^2.$
The pulloff force drops rapidly for $h_\text{rms}/\delta_c >1$ and 
is extremely small for $h_\text{rms}/\delta_c >3$.
In contrast to our results, no clear transition to nonadhesive
behavior with area proportional to load
was discussed.

These traditional theories expressed the pulloff force as a ratio
to the maximum force $N P_c$ where $N$ is the number of spherical
asperities and $P_c = 3\pi w R/2$ the pulloff force for each
in the JKR limit.
 From statistical studies of rough surfaces,\cite{Nayak,Archard:1974,Maugis:1996_2}
 $N \sim 0.05 A_0 / R h_\text{rms}$.
This gives $N P_c \sim w A_0 / 4 h_\text{rms}$, with no
explicit dependence on $R$.

\begin{figure}
  \begin{center}
    \includegraphics[width=12cm]{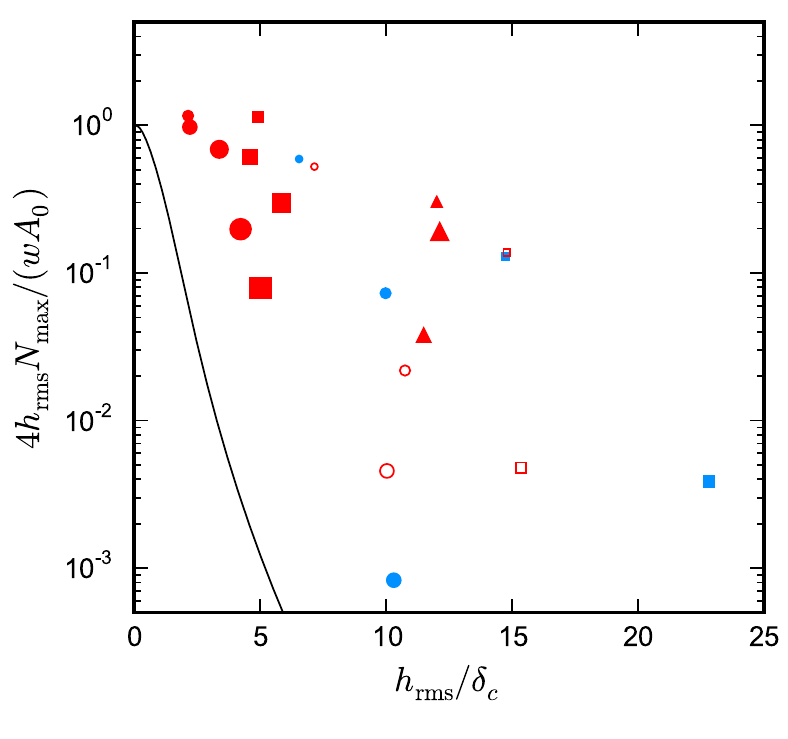}
    \caption{\label{pulloff_gw}
    {\bf Comparison with Greenwood-Williamson-type theories.}
    Maximum force $N_\mathrm{max}$ upon approach of the two surfaces plotted against the adhesion parameter $h_\text{rms}/\delta_c$ with $\delta_c=3/4(\pi^2 w^2 R/(E^*)^2)^{1/3}$ identified by Fuller \& Tabor.~\cite{Fuller:1975p327_2}
    The force is plotted in units of $NP_c$ where $P_c=3\pi w R/2$ is the Johnson-Kendall-Roberts~\cite{Johnson:1971p301} pulloff force for a single sphere and $N$ is the number of asperities.
    We use the rms curvature to express $1/R=h_\text{rms}''/2$ and additionally make use the frequently quoted relationship $R h_\text{rms} N/A_0=0.05$.~\cite{Archard:1974}
    Clearly, there is little correlation between adhesion parameter and maximum force.
Results are shown for $H=0.3$ (triangles), 0.5 (squares) 
and 0.8 (circles)
and $\ell_a/a_0 = 0.005$ (blue) and 0.05 (red).
Closed and open symbols are for $h_{\rm rms}'=0.1$ and 0.3, respectively.
The symbol size increases as $\lambda_s/a_0$ increases from 4 to 64 
in powers of 2.
      }
  \end{center}
\end{figure}

Figure \ref{pulloff_gw} shows our results for the adhesive force as a function
of $h_\text{rms}/\delta_c$ and the prediction of Fuller and Tabor,
which is very close to the expressions obtained by Maugis.
Note that we plot the maximum force $N_\text{max}$ obtained as surfaces
are brought together, because the pulloff force is not unique and
depends on the peak loading pressure.
However the pulloff force is always larger than $N_\text{max}$
which would move the numerical data even farther from the theoretical
prediction.

It is clear that traditional theories are both quantitatively and
qualitatively inconsistent with numerical solutions of the model
they were derived for.
The predicted pulloff force falls orders of magnitude below the
numerical results,
and systems with the same value of $h_\text{rms}/\delta_c$
have pulloff forces that vary by almost three orders of magnitude.
As noted in the main text, the numerical results depend only
on the rms slope and curvature which are predominantly determined
by small wavelength roughness.
The numerical results do not vary with the long wavelength cutoff of roughness
$\lambda_L$, while $h_\text{rms} \sim \lambda_L^H$ changes more than an
order of magnitude.

Fuller and Tabor did not actually use their model to fit their data.
They noted that the number of asperities should vary as $1/R h_\text{rms}$ but then dropped
this dependence.
Instead, they normalized by the pulloff force for smooth surfaces of the
same chemistry.
The data were then collapsed by fitting $\delta_c$ to find an effective
radius rather than obtaining $R$ from the actual surface.
The resulting radius was about 50$\mu$m, which is much larger than the
smallest asperities on typical surfaces.
This approach of rescaling both axes to match the theoretical prediction
has been typical of subsequent work and masks quantitative errors in the
theory.
Note that normalizing our numerical data by the smooth surface result,
$w A_0/\Delta r$, introduces a parameter that is not present in past
theories and does not improve the correlation between
pulloff force and $h_\text{rms}/\delta_c$.


\newpage

\end{document}